\documentclass{article}

\usepackage{graphicx}

\usepackage{amsmath}
\usepackage{times,txfonts}

\begin{document}

\title{Multi-Dimensional Nonsystematic Reed-Solomon Codes}
\author{Akira Shiozaki \thanks{Emeritus professor, Osaka Prefecture University, Japan. \ E-mail:\ shiozaki.akira@gmail.com}}
\maketitle

\begin{abstract}
This paper proposes a new class of multi-dimensional nonsystematic Reed-Solomon codes that are constructed based on 
the multi-dimensional Fourier transform over a finite field. 
The proposed codes are the extension of the nonsystematic Reed-Solomon codes to multi-dimension. 
This paper also discusses the performance of the multi-dimensional nonsystematic Reed-Solomon codes. 
\end{abstract}

\vspace{3pt}


\section{Introduction}

Many error-correcting-codes \cite{mac},\cite{blahut} have been developed to enhance the reliability of 
data transmission systems and memory systems. 
One class of superior error-correcting-codes is the Reed-Solomon codes that are maximum-distance codes. 
The nonsystematic Reed-Solomon codes \cite{reed} are constructed based on the one-dimensional Fourier transforms 
over a finite field. 
The code length of the nonsystematic Reed-Solomon codes over a finite field $GF(q)$ is $q$, 
while the code length of the systematic and cyclic Reed-Solomon codes is $q-1$. 

The author presented the two-dimensional nonsystematic Reed-Solomon codes based on two-dimensional 
Fourier transform \cite{shiozaki1} and showed the extension of the codes to multi-dimensional codes \cite{shiozaki2}. 
On the other hand, Shen, et al. \cite{shen} presented the multidimensional extension of the Reed-Solomon codes 
using a location set contained in a multidimensional affine or projective space over a finite field. 
But they described only the two-dimensional extension concretely. 

This paper proposes a new class of multi-dimensional nonsystematic Reed-Solomon codes that are constructed based on 
the multi-dimensional Fourier transform over a finite field. 
The proposed codes are the extension of the nonsystematic Reed-Solomon codes to multi-dimension, 
and are the developments of the codes in \cite{shiozaki2}. 
The code length of the $n$-dimensional nonsystematic Reed-Solomon codes over a finite field $GF(q)$ is $q^n$. 
This paper also discusses the performance of the multi-dimensional nonsystematic Reed-Solomon codes. 

\section{$2$-dimensional Reed-Solomon codes}

Firstly, we consider the following codes based on $2$-dimensional Fourier transform. 

Let $a_{ij}\ (0 \leq i \leq K_j; 0 \leq j \leq L)$ be any elements of a finite field $GF(q)$ 
and let $f(x_1,x_2)$ be a polynomial of two variables whose coefficients are $a_{ij}$:
\begin{align}
 f(x_1,x_2) &= \sum_{j=0}^L \left( \sum_{i=0}^{K_j} a_{ij} x_1^i \right) x_2^j  \notag \\
              &= \sum_{j=0}^L f_j(x_1) x_2^j \ \ \ \ \ \ \ (L \leq q-1)
\end{align}
\begin{equation}
 f_j(x_1) = \sum_{i=0}^{K_j} a_{ij} x_1^i \ \ \ \ \ \ \ (K_j \leq q-1)
\end{equation}

We consider the code whose codeword consists of $q^2$ elements $\{f(\beta_k,\beta_l)\}\ (k=0,1,\cdots,q-1; l=0,1,\cdots,q-1)$, 
where $\beta_k$ and $\beta_l$ are any elements of $GF(q)$. 
The transformation of the information symbols $\{a_{ij}\}$ to a codeword $\{f(\beta_k,\beta_l)\}$ is 
the two-dimensional Fourier transform over $GF(q)$, 
and so the code is the two-dimensional extension of a nonsystematic Reed-Solomon code. 
The code length $N$ is $N=q^2$. 

When $f_j(x_1) \not= 0$, the number of $\beta_k\ (0 \leq k \leq q-1)$ such that $f_j(\beta_k) \not= 0$ is at least $q-K_j$, 
because the number of the roots of $f_j(x_1)$ is at most $K_j$. 

A nonzero codeword has at least one $f_j(x_1)\ (0 \leq j \leq L)$ such that $f_j(x_1) \not= 0$. 
Now let $m$ be the maximum $j$ of the nonzero $f_j(x_1)$, that is, 
$f_m(x_1) \not= 0,\ f_{m+1}(x_1)=f_{m+2}(x_1)= \cdots =f_L (x_1)=0$. 
The number of $\beta_k$ such that $f_m(\beta_k) \not= 0$ is at least $q-K_m$. 
For an element $\beta_k$ such that $f_j(\beta_k) \not= 0\ (0 \leq j \leq m)$, 
the number of $\beta_l$ such that $f(\beta_k,\beta_l) \not= 0$ is at least $q-m$ 
because the number of the roots of 
\begin{equation}
 f(\beta_k, x_2) =  \sum_{j=0}^m f_j(\beta_k) x_2^j 
\end{equation}

\noindent is at most $m$. 
Therefore the number of the pairs $(\beta_k,\beta_l)$ such that $f(\beta_k,\beta_l) \not= 0$ is at least 
\begin{equation}
 \min_{0\leq m \leq L} [(q-K_m)(q-m)]
\label{dmin2}
\end{equation}

\noindent and it is equal to the minimum distance $d_{min}$ of the code. 
From Eq.(\ref{dmin2}), $K_m=q-\lceil \frac{d_{min}}{q-m} \rceil$\footnote{$\lceil x \rceil$ denotes 
the minimum integer not less than $x$} because $q-K_m \ (m=0,1,\cdots,L)$ must be $\lceil \frac{d_{min}}{q-m} \rceil$. 
$L$ should be determined as the maximum integer such that $K_L=q-\lceil \frac{d_{min}}{q-L} \rceil \geq 0$. 
Then the number of the information symbols $K$ is 
\begin{equation}
 K = \sum_{m=0}^L (K_m+1) = \sum_{m=0}^L \left( q + 1 - \lceil \frac{d_{min}}{q-m} \rceil \right) 
\end{equation}

\noindent and the number of the check symbols $N-K=q^2-K$ is 
\begin{equation}
 N-K = \sum_{m=0}^L \left( \lceil \frac{d_{min}}{q-m} \rceil - 1 \right) + q(q-L-1).
\end{equation}

The above statement is summarized in the following theorem:

\vspace{3pt}
[Theorem 1]\ Let $a_{ij}\ (0 \leq i \leq K_j; 0 \leq j \leq L)$ be any elements of a finite field $GF(q)$, 
where $K_j$ is $K_j=q-\lceil \frac{d_{min}}{q-j} \rceil$ and $L$ is the maximum integer 
such that $K_L=q-\lceil \frac{d_{min}}{q-L} \rceil \geq 0$. 

For a polynomial of two variables such that 
\begin{equation}
 f(x_1,x_2) = \sum_{j=0}^L \sum_{i=0}^{K_j} a_{ij} x_1^i x_2^j \ , 
\end{equation}
\noindent the code whose codeword consists of $q^2$ elements $\{f(\beta_k,\beta_l)\}\ (k=0,1,\cdots,q-1; l=0,1,\cdots,q-1)$ 
is a linear code with minimum distance $d_{min}$, 
where $\beta_k$ and $\beta_l$ are the elements of $GF(q)$. 
$\Box$
\vspace{3pt}

\begin{figure}[h]
  \begin{center}
    \includegraphics[width=4.5cm,clip]{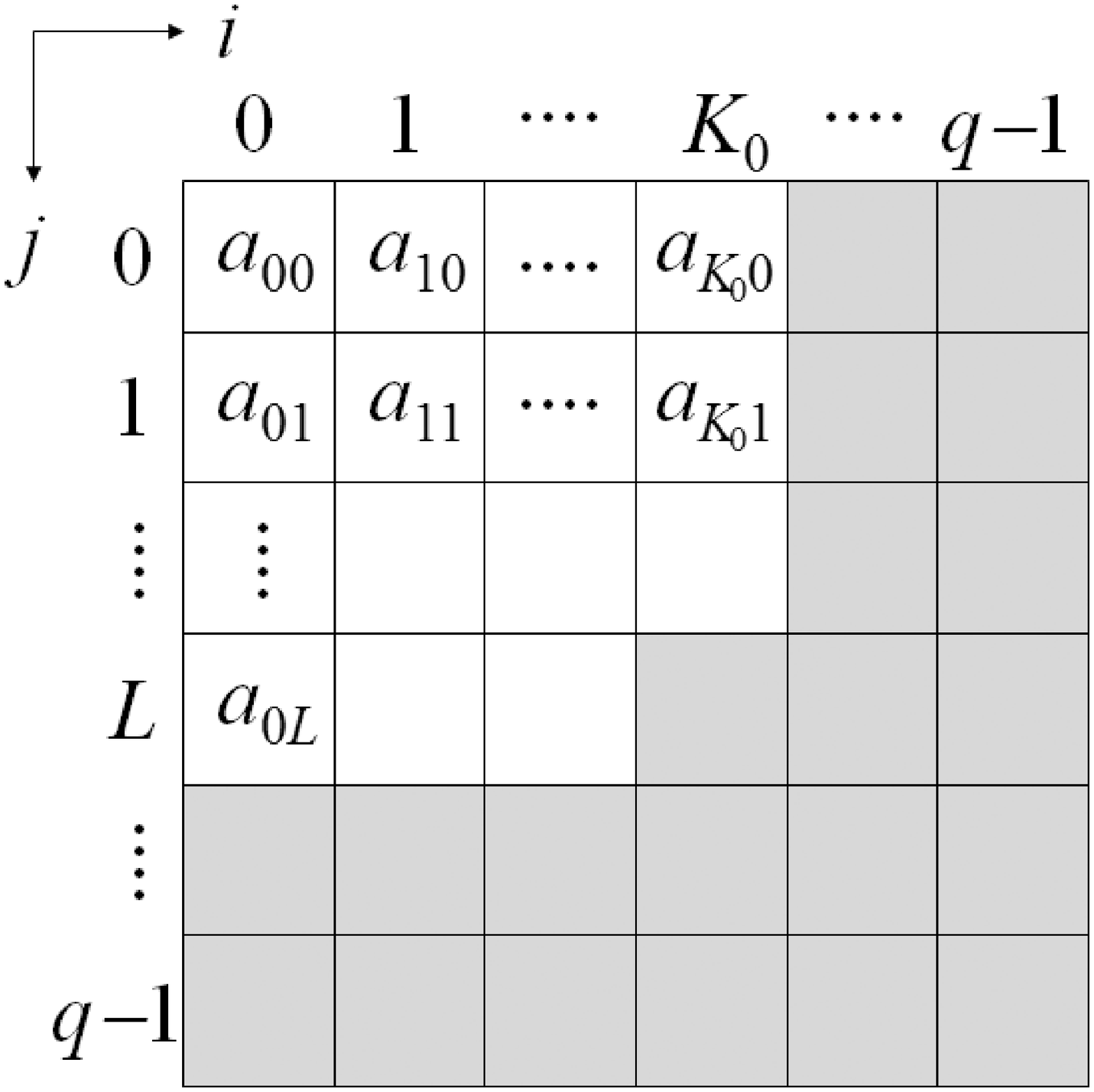}
    \caption[]{$2$-dimensional Reed-Solomon codes}
    \label{2D_RScode}
  \end{center}
\end{figure}

Figure \ref{2D_RScode} shows the example of a $2$-dimensional Reed-Solomon code. 
Table \ref{information} shows the distribution of $K_m$ in case of $q=5$. 

\begin{table}[h]
 \begin{center}
 \caption{{\footnotesize Number of information symbols of $2$-dimensional Reed-Solomon code ($q=5$)}}
 \begin{tabular}{c|c|c||c|c|c||c|c|c||c|c|c} \hline
 $d_{min}$ & $m$ & $K_m$ & $d_{min}$ & $m$ & $K_m$ & $d_{min}$ & $m$ & $K_m$ & $d_{min}$ & $m$ & $K_m$ \\ \hline
 $3$        & $0$  & $4$     & $4$        & $0$  & $4$   & $5$         & $0$  & $4$     & $6$        & $0$  & $3$     \\ 
              & $1$  & $4$     &              & $1$  & $4$    &               & $1$  & $3$     &              & $1$  & $3$     \\ 
              & $2$  & $4$     &              & $2$  & $3$    &               & $2$  & $3$     &              & $2$  & $3$     \\ 
              & $3$  & $3$     &              & $3$  & $3$    &               & $3$  & $2$     &              & $3$  & $2$     \\ 
              & $4$  & $2$     &              & $4$  & $1$    &               & $4$  & $0$     &              &        &            \\ \cline{2-3}\cline{5-6}\cline{8-9}\cline{11-12}
 & \multicolumn{2}{c||}{$K = 22$} & & \multicolumn{2}{c||}{$K = 20$} & & \multicolumn{2}{c||}{$K = 17$} & & \multicolumn{2}{c}{$K = 15$} \\ \hline \hline
 $d_{min}$ & $m$ & $K_m$ & $d_{min}$ & $m$ & $K_m$ & $d_{min}$ & $m$ & $K_m$ & $d_{min}$ & $m$ & $K_m$ \\ \hline
 $7$        & $0$  & $3$     & $8$        & $0$  & $3$   & $9$         & $0$  & $3$     & $10$      & $0$  & $3$     \\ 
              & $1$  & $3$     &              & $1$  & $3$    &               & $1$  & $2$     &              & $1$  & $2$     \\ 
              & $2$  & $2$     &              & $2$  & $2$    &               & $2$  & $2$     &              & $2$  & $1$     \\ 
              & $3$  & $1$     &              & $3$  & $1$    &               & $3$  & $0$     &              & $3$  & $0$     \\ \cline{2-3}\cline{5-6}\cline{8-9}\cline{11-12}
 & \multicolumn{2}{c||}{$K = 13$} & & \multicolumn{2}{c||}{$K = 13$} & & \multicolumn{2}{c||}{$K = 11$} & & \multicolumn{2}{c}{$K = 10$} \\ \hline
\end{tabular}
 \label{information}
 \end{center}
\end{table}

The number of the information symbols $K$ is 
\begin{align}
 K &= \sum_{m=0}^L \left( q + 1 - \lceil \frac{d_{min}}{q-m} \rceil \right)  \notag \\
    &\geq  \sum_{m=0}^L \left( q - \frac{d_{min}}{q-m} \right)  \ \ \ \ \ \ \ \ \ \ \ \ \ \ \left(\textrm{because}\ \lceil \frac{d_{min}}{q-m} \rceil \leq \frac{d_{min}}{q-m} + 1 \right)  \notag \\
    &= q(L+1) - \sum_{m=0}^L \frac{d_{min}}{q-m}  \notag \\
    &> q \left( q - \frac{d_{min}}{q} \right) - d_{min} \sum_{m=0}^L \frac{1}{q-m} \ \ \ \ \ \ \ \ \ \ \ \ \ \ \left( L > q-\frac{d_{min}}{q}-1\ \textrm{because}\ q \geq \lceil \frac{d_{min}}{q-L} \rceil \right)  \notag \\
    &= q^2 - d_{min} - d_{min} \sum_{m=0}^L \frac{1}{q-m}  \notag \\
    &> q^2 - d_{min} - d_{min} \sum_{m=0}^{\lfloor q-\frac{d_{min}}{q} \rfloor} \frac{1}{q-m}  \ \ \ \ \ \ \ \ \ \ \ \ \ \ \left( L \leq q - \frac{d_{min}}{q} \right).  
\end{align}
\noindent So 
\begin{equation}
 \frac{K}{N} > 1 - \frac{d_{min}}{N} - \frac{d_{min}}{N} \sum_{m=0}^{\lfloor q-\frac{d_{min}}{q} \rfloor} \frac{1}{q-m} 
\label{kn}
\end{equation}

\begin{figure}[h]
  \begin{center}
    \includegraphics[width=7.8cm,clip]{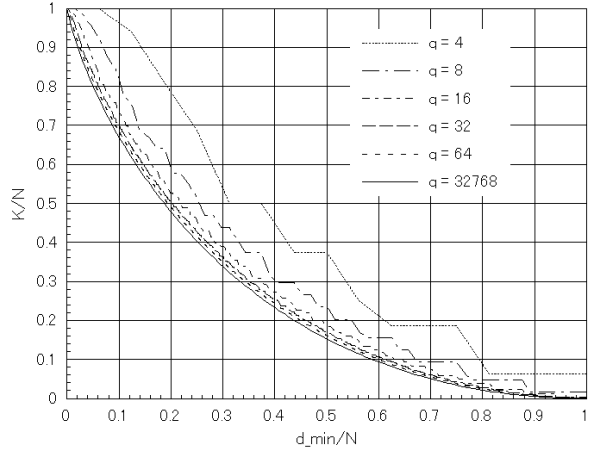}
   \caption[]{Relation between $d_{min}/N$ and $K/N$ ($2$-dimensional)}
    \label{dn_kn}
  \end{center}
\end{figure}

\vspace{3pt}
\noindent Figure \ref{dn_kn} shows the relation between $d_{min}/N$ and $K/N$. 

\section{$3$-dimensional Reed-Solomon codes}

We extend the discussion in the preceding chapter to $3$-dimensional Fourier transform over a finite field. 

Let $a_{i_1i_2 i_3}\ (0 \leq i_1 \leq K_{i_2 i_3} ; \ 0 \leq i_2 \leq L_{i_3}; 0 \leq i_3 \leq L)$ be 
any elements of a finite field $GF(q)$, 
and let $f(x_1,x_2,x_3)$ be a polynomial of three variables whose coefficients are $a_{i_1 i_2 i_3}$:
\begin{align}
 f(x_1,x_2,x_3) &= \sum_{i_3=0}^{L} \sum_{i_2=0}^{L_{i_3}} \left( \sum_{i_1=0}^{K_{i_2 i_3}} a_{i_1 i_2 i_3} x_1^{i_1} \right) x_2^{i_2} x_3^{i_3}  \notag \\ 
              &= \sum_{i_3=0}^L \sum_{i_2=0}^{L_{i_3}} f_{i_2 i_3} (x_1)\ x_2^{i_2} x_3^{i_3}  
\end{align}
\begin{equation}
 f_{i_2 i_3} (x_1) = \sum_{i_1=0}^{K_{i_2 i_3}} a_{i_1 i_2 i_3} x_1^{i_1} \ \ \ \ \ \ \ (K_{i_2 i_3} \leq q-1)
\end{equation}
We consider the code whose codeword consists of $q^3$ elements $\{f(\beta_{k_1},\beta_{k_2},\beta_{k_3})\}\ (k_j=0,1,\cdots,q-1)$, 
where $\beta_{k_j}\ (j=1,2,3)$ are any elements of $GF(q)$. 
The transformation of the information symbols $\{a_{i_1 i_2 i_3}\}$ to a codeword $\{f(\beta_{k_1},\beta_{k_2},\beta_{k_3})\}$ is 
the three-dimensional Fourier transform over $GF(q)$, 
and so the code is the three-dimensional extension of a nonsystematic Reed-Solomon code. 
The code length $N$ is $N=q^3$. 

When $f_{i_2 i_3} (x_1) \not= 0$, the number of $\beta_{k_1}\ (0 \leq k_1 \leq q-1)$ such that $f_{i_2 i_3} 
(\beta_{k_1}) \not= 0$ is at least $q-K_{i_2 i_3}$, 
because the number of the roots of $f_{i_2 i_3} (x_1)$ is at most $K_{i_2 i_3}$. 

Now let $m_2$ be the maximum $i_2$ of  the nonzero $f_{i_2 i_3} (x_1)$ and let $m_3$ be the maximum $i_3$ of 
the nonzero $f_{i_2 i_3} (x_1)$. 
Then let $K_{m_2 m_3}$ be the maximum $i_1$ in this case. 

For the equations 
\begin{align}
 f(\beta_{k_1}, x_2, x_3) &= \sum_{i_3=0}^{m_3} \left( \sum_{i_2=0}^{m_2} f_{i_2 i_3} (\beta_{k_1}) x_2^{i_2} \right) x_3^{i_3}  \notag \\
                          &= \sum_{i_3=0}^{m_3} f_{i_3} (\beta_{k_1}, x_2)\ x_3^{i_3}  
\end{align}
and
\begin{equation}
 f_{i_3} (\beta_{k_1}, x_2) = \sum_{i_2=0}^{m_2} f_{i_2 i_3} (\beta_{k_1}) x_2^{i_2}\ ,
\end{equation}
\noindent the number of $\beta_{k_2}$ such that $f_{i_3} (\beta_{k_1}, \beta_{k_2}) \not= 0$ is at least $q-m_2$ 
because the number of the roots of $f_{i_3} (\beta_{k_1}, x_2)$ is at most $m_2$. 
For $\beta_{k_2}$ such that $f_{i_3} (\beta_{k_1}, \beta_{k_2}) \not= 0$, 
the number of $\beta_{k_3}$ such that $f(\beta_{k_1},\beta_{k_2},\beta_{k_3}) \not= 0$ is at least $q-m_3$ 
because the number of the roots of 
\begin{equation}
f(\beta_{k_1}, \beta_{k_2}, x_3) = \sum_{i_3=0}^{m_3} f_{i_3} (\beta_{k_1}, \beta_{k_2})\ x_3^{i_3}  
\end{equation}
\noindent is at most $m_3$. 
Therefore the number of the three-tuples $(\beta_{k_1},\beta_{k_2},\beta_{k_3})$ 
such that $f(\beta_{k_1},$ $\beta_{k_2},\beta_{k_3})$ $\not= 0$ is at least 
\begin{equation}
 \min_{\substack{{0\leq m_3 \leq L}\\ {0\leq m_2 \leq L_{m_3}}}} [(q-K_{m_2 m_3})(q-m_2)(q-m_3)]\ ,
\label{dmin3}
\end{equation}
and it is equal to the minimum distance $d_{min}$ of the code. 

From Eq.(\ref{dmin3}), $K_{m_2 m_3} = q-\lceil \frac{d_{min}}{(q-m_2)(q-m_3)} \rceil$ 
because $q-K_{m_2 m_3} \ (m_2=0,1,\cdots,L_{m_3}; m_3=0,1,\cdots,L)$ must be $\lceil \frac{d_{min}}{(q-m_2)(q-m_3)} \rceil$. 
$L_{m_3}$ and $L$ should be respectively determined as the maximum $m_2$ and the maximum $m_3$ 
such that $K_{m_2 m_3}=q-\lceil \frac{d_{min}}{(q-m_2)(q-m_3)} \rceil \geq 0$. 
Then the number of the information symbols $K$ is 
\begin{equation}
 K = \sum_{m_3=0}^L \sum_{m_2=0}^{L_{m_3}} (K_{m_2 m_3}+1) = \sum_{m_3=0}^L \sum_{m_2=0}^{L_{m_3}} \left( q + 1 - \lceil \frac{d_{min}}{(q-m_2)(q-m_3)} \rceil \right) 
\end{equation}
\noindent and the number of the check symbols $N-K=q^3-K$ is 
\begin{equation}
 N-K = \sum_{m_3=0}^L \sum_{m_2=0}^{L_{m_3}} \left( \lceil \frac{d_{min}}{(q-m_2)(q-m_3)} \rceil - 1 \right) + q^3 -q(L+1)(L_{m_3}+1)\  .
\end{equation}
The above statement is summarized in the following theorem:

\vspace{3pt}
[Theorem 2]\ Let $a_{i_1i_2 i_3}\ (0 \leq i_1 \leq K_{i_2 i_3} ; \ 0 \leq i_2 \leq L_{i_3};\ 0 \leq i_3 \leq L)$ be 
any elements of a finite field $GF(q)$, 
where $K_{i_2 i_3} = q-\lceil \frac{d_{min}}{(q-i_2)(q-i_3)} \rceil$ and $L_{i_3}$ and $L$ are the maximum integers 
such that $K_{L_{i_3} L}=q-\lceil \frac{d_{min}}{(q-L_{i_3})(q-L)} \rceil \geq 0$. 

For a polynomial of three variables such that 
\begin{equation}
 f(x_1,x_2,x_3) = \sum_{i_3=0}^{L} \sum_{i_2=0}^{L_{i_3}} \sum_{i_1=0}^{K_{i_2 i_3}} a_{i_1 i_2 i_3} x_1^{i_1} x_2^{i_2} x_3^{i_3}\ , 
\end{equation}
\noindent the code whose codeword consists of $q^3$ elements $\{f(\beta_{k_1},\beta_{k_2},\beta_{k_3})\}\ (k_l=0,1,\cdots,q-1;\ l=1,2,3)$ 
is a linear code with minimum distance $d_{min}$, 
where $\beta_{k_1},\beta_{k_2},\beta_{k_3}$ are the elements of $GF(q)$. 
$\Box$ 

\section{$n$-dimensional Reed-Solomon codes} 

We extend the discussion in the preceding chapter to $n$-dimensional Fourier transform over a finite field. 

Let $a_{i_1i_2\cdots i_n}\ (0 \leq i_1 \leq K_{i_2 i_3 \cdots i_n} ;\ 0 \leq i_j \leq L_{i_{j+1} i_{j+2} \cdots i_n}; j=2,3,\cdots,n-1;\ 0 \leq i_n \leq L)$ 
be any elements of a finite field $GF(q)$, 
and let a polynomial of $n$ variables whose coefficients are $a_{i_1i_2\cdots i_n}$: 
\begin{align}
 f(x_1,x_2,\cdots,x_n) &= \sum_{i_n=0}^L \sum_{i_{n-1}=0}^{L_{i_n}} \cdots \sum_{i_2=0}^{L_{i_3 i_4 \cdots i_n}} \left( \sum_{i_1=0}^{K_{i_2 i_3 \cdots i_n}} a_{i_1i_2 \cdots i_n} x_1^{i_1} \right) x_2^{i_2} \cdots x_n^{i_n}  \notag \\ 
              &= \sum_{i_n=0}^L \sum_{i_{n-1}=0}^{L_{i_n}} \cdots \sum_{i_2=0}^{L_{i_3 i_4 \cdots i_n}} f_{i_2 i_3 \cdots i_n} (x_1)\ x_2^{i_2} \cdots x_n^{i_n}  
\end{align}
\begin{equation}
 f_{i_2 i_3 \cdots i_n} (x_1) = \sum_{i_1=0}^{K_{i_2 i_3 \cdots i_n}} a_{i_1 i_2 \cdots i_n} x_1^{i_1} \ \ \ \ \ \ \ (K_{i_2 i_3 \cdots i_n} \leq q-1)
\end{equation}
We consider the code whose codeword consists of $q^n$ elements $\{f(\beta_{k_1},\beta_{k_2},\cdots,\beta_{k_n})\}\ $ $(k_j=0,1,\cdots,q-1)$, 
where $\beta_{k_j}\ (j=1,2,\cdots,n)$ are any elements of $GF(q)$. 
The transformation of the information symbols $\{a_{i_1i_2\cdots i_n}\}$ to a codeword $\{f(\beta_{k_1},\beta_{k_2},\cdots,\beta_{k_n})\}$ is 
the $n$-dimensional Fourier transform over $GF(q)$, 
and so the code is the $n$-dimensional extension of a nonsystematic Reed-Solomon code. 
The code length $N$ is $N=q^n$. 

From the discussion in the preceding chapter, the number of $n$-tuples $(\beta_{k_1},$ $\beta_{k_2},$ $\cdots,$ $\beta_{k_n})$ such that $f(\beta_{k_1},$ $\beta_{k_2},$ $\cdots,$ $\beta_{k_n}) \not= 0$ 
is at least 
\begin{equation}
 \min_{\substack{{0\leq m_n \leq L}\\ 0\leq m_j \leq L_{m_{j+1} m_{j+2} \cdots m_n}\ (j=2,3,\cdots,n-1)}} [(q-K_{m_2 m_3 \cdots m_n})(q-m_2)(q-m_3)\cdots(q-m_n) ]
\label{dmin_n}
\end{equation}
\noindent ant it is equal to the minimum distance $d_{min}$ of the code. 

From Eq.(\ref{dmin_n}), $K_{m_2 m_3 \cdots m_n} = q-\lceil \frac{d_{min}}{(q-m_2)(q-m_3)\cdots(q-m_n)} \rceil$ 
because $q-K_{m_2 m_3 \cdots m_n} \ (m_j=0,1,\cdots,L_{m_{j+1} m_{j+2} \cdots m_n};$\ $j=2,3,\cdots,n-1;\ m_n=0,1,\cdots,L)$ 
must be $\lceil \frac{d_{min}}{(q-m_2)(q-m_3)\cdots(q-m_n)} \rceil$. 
$L_{m_3 m_4 \cdots m_n}, L_{m_4 m_5 \cdots m_n}, \cdots, L_{m_n}$ and $L$ should be respectively determined as the maximum $m_2, m_3, \cdots, m_n$ 
such that $K_{m_2 m_3 \cdots m_n}$$=q-\lceil \frac{d_{min}}{(q-m_2)(q-m_3)\cdots(q-m_n)} \rceil \geq 0$. 

The above statement is summarized in the following theorem:

\vspace{3pt}

[Theorem 3]\ Let $a_{i_1i_2\cdots i_n}\ (0 \leq i_1 \leq K_{i_2 i_3 \cdots i_n} ;\ 0 \leq i_j \leq L_{i_{j+1} i_{j+2} \cdots i_n};\ $ $j=2,3,\cdots,n-1;\ 0 \leq i_n \leq L)$ be 
any elements of a finite field $GF(q)$, 
where $K_{i_2 i_3 \cdots i_n} =$ $q - \lceil \frac{d_{min}}{(q-i_2)(q-i_3)\cdots(q-i_n)} \rceil$ and $L_{i_3 i_4 \cdots i_n},\ $ $L_{i_4 i_5 \cdots i_n}, \cdots, L_{i_n}$ and $L$ 
are the maximum integers such that $K_{i_2 i_3 \cdots i_n}$ $=q-\lceil \frac{d_{min}}{(q-i_2)(q-i_3)\cdots(q-i_n)} \rceil \geq 0$. 

For a polynomial of $n$ variables such that 
\begin{equation}
 f(x_1,x_2,\cdots,x_n) = \sum_{i_n=0}^L \sum_{i_{n-1}=0}^{L_{i_n}} \cdots \sum_{i_2=0}^{L_{i_3 i_4 \cdots i_n}} \sum_{i_1=0}^{K_{i_2 i_3 \cdots i_n}} a_{i_1i_2 \cdots i_n} x_1^{i_1} x_2^{i_2} \cdots x_n^{i_n}
\end{equation}
\noindent the code whose codeword consists of $q^n$ elements $\{f(\beta_{k_1},\beta_{k_2},\cdots,\beta_{k_n})\}$\ $(k_l=0,1,\cdots,q-1;\ l=1,2,\cdots,n)$ 
is a linear code with minimum distance $d_{min}$, where $\beta_{k_1},\beta_{k_2},\cdots,\beta_{k_n}$ are the elements of $GF(q)$. 
$\Box$ 

\vspace{3pt}

The number of the information symbols $K$ is 
\begin{equation}
 K = \sum_{i_n=0}^L \sum_{i_{n-1}=0}^{L_{i_n}} \cdots \sum_{i_2=0}^{L_{i_3 i_4 \cdots i_n}} \left( K_{i_2 i_3 \cdots i_n} + 1 \right) = \sum_{i_n=0}^L \sum_{i_{n-1}=0}^{L_{i_n}} \cdots \sum_{i_2=0}^{L_{i_3 i_4 \cdots i_n}} \left( q + 1 - \lceil \frac{d_{min}}{\prod^{n}_{j=2} (q-i_j)} \rceil \right) 
\end{equation}
\noindent and the number of the check symbols $N-K=q^n-K$ is 

\begin{equation}
 N-K = \sum_{i_n=0}^L \sum_{i_{n-1}=0}^{L_{i_n}} \cdots \sum_{i_2=0}^{L_{i_3 i_4 \cdots i_n}} \left( \lceil \frac{d_{min}}{\prod^{n}_{j=2} (q-i_j)} \rceil - 1 \right) + q^n -q(L+1)\prod^{n-1}_{j=2}(L_{i_{j+1} i_{j+2} \cdots i_n}+1)\ .
\end{equation}
When $L=q-1$ and $L_{i_{j+1} i_{j+2} \cdots i_n}=q-1\ (j=2,3,\cdots,n-1)$, that is, $d_{min}\leq q$, the number of the check symbols $N-K$ is 

\begin{align}
 N-K &= \sum_{i_n=0}^{q-1} \sum_{i_{n-1}=0}^{q-1} \cdots \sum_{i_2=0}^{q-1} \left( \lceil \frac{d_{min}}{\prod^{n}_{j=2} (q-i_j)} \rceil - 1 \right) \notag \\
     &= \sum_{i_n=q-d_{min}+1}^{q-1} \sum_{i_{n-1}=q-d_{min}+1}^{q-1} \cdots \sum_{i_2=q-d_{min}+1}^{q-1} \left( \lceil \frac{d_{min}}{\prod^{n}_{j=2} (q-i_j)} \rceil - 1 \right) \notag \\
     &= \sum_{i_n=1}^{d_{min}-1} \sum_{i_{n-1}=1}^{d_{min}-1} \cdots \sum_{i_2=1}^{d_{min}-1} \left( \lceil \frac{d_{min}}{\prod^{n}_{j=2} i_j} \rceil - 1 \right)\ .
\end{align}
\noindent The number of the check symbols $N-K$ has no relation to the number $q$ of the elements of a finite field $GF(q)$ 
and is determined by only the minimum distance $d_{min}$. 
Table \ref{check} shows the number of the check symbols when $d_{min} \leq q$. 

\begin{table}[h]
 \begin{center}
 \caption{Number of check symbols $(N-K)$ when $d_{min} \leq q$}
 \begin{tabular}{c|c|c|c|c} \hline
 & \multicolumn{4}{c}{$N-K$} \\ \hline
 $d_{min}$ & $n=2$ & $n=3$ & $n=4$ & $n=5$ \\ \hline
 $2$        & $1$     & $1$    & $1$     & $1$ \\ 
 $3$        & $3$     & $4$    & $5$     & $6$ \\ 
 $4$        & $5$     & $7$    & $9$     & $11$ \\ 
 $5$        & $8$     & $13$  & $19$   & $26$ \\ 
 $6$        & $10$   & $16$  & $23$   & $31$ \\ 
 $7$        & $14$   & $25$  & $39$   & $56$ \\ 
 $8$        & $16$   & $28$  & $43$   & $61$ \\ 
 $9$        & $20$   & $38$  & $63$   & $96$ \\ 
 $10$      & $23$   & $44$  & $73$   & $111$ \\ 
 $11$      & $27$   & $53$  & $89$   & $136$ \\ 
 $12$      & $29$   & $56$  & $93$   & $141$ \\ 
 $13$      & $35$   & $74$  & $133$  & $216$ \\ 
 $14$      & $37$   & $77$  & $137$  & $221$ \\ 
 $15$      & $41$   & $86$  & $153$  & $246$ \\ 
 $16$      & $45$   & $95$  & $169$  & $271$ \\ \hline
 \end{tabular}
 \label{check}
 \end{center}
\end{table}

\section{Performance}

\subsection{Comparison between $2$-dimensional Reed-Solomon codes and product codes}

The product code of a $(n_1,k_1,d_1)$ linear code and a $(n_2,k_2,d_2)$ linear code is a $(N,K,d_{min})=(n_1n_2,k_1k_2,d_1d_2)$ 
linear code. 
When two linear codes are the same $(n,k,d)$ Reed-Solomon codes over $GF(q)$, the number of the check symbols of 
the product code is 
\begin{equation}
 N - K = (d-1)(2n-d+1)\ . 
\end{equation}
\noindent Then the relation between $\frac{d_{min}}{N}$ and $\frac{K}{N}$ is 
\begin{equation}
 1 - \frac{K}{N} = \left( \sqrt{\frac{d_{min}}{N}}-\frac{1}{q} \right) \left( 2-\sqrt{\frac{d_{min}}{N}}+\frac{1}{q} \right) 
\end{equation}
\noindent when $n=q$. 

\begin{figure}[h]
  \begin{center}
    \includegraphics[width=7.8cm,clip]{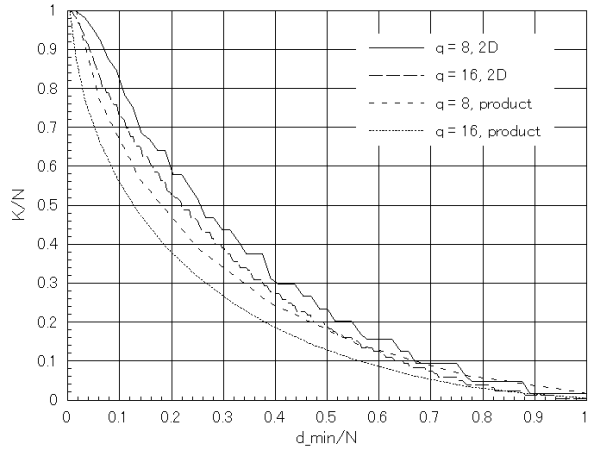}
    \caption[]{Performances of $2$-dimensional codes and product codes}
    \label{product_perform}
  \end{center}
\end{figure}

Figure \ref{product_perform} shows the relations between $\frac{d_{min}}{N}$ and $\frac{K}{N}$ of the $2$-dimensional Reed-Solomon codes and the product codes 
when $q=8$ and $q=16$. 
As shown in Fig.\ref{product_perform}, the performance of the $2$-dimensional codes is higher than that of the product codes. 

\subsection{Relation between dimension and performance}

\begin{figure}[h]
  \begin{center}
    \includegraphics[width=7.8cm,clip]{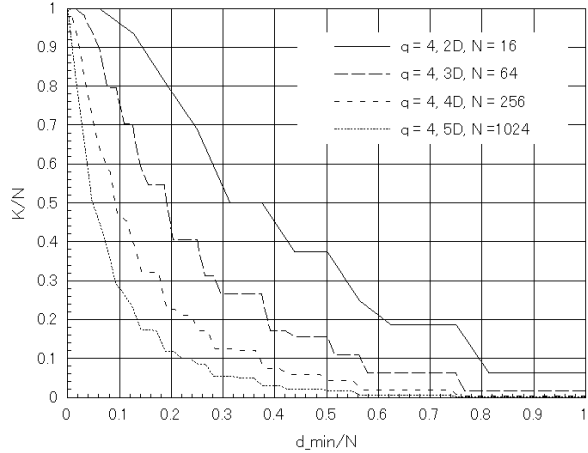}
    \caption[]{Relation between $d_{min}/N$ and $K/N$\ $(q=4)$}
    \label{q4_dn_kn}
  \end{center}
\end{figure}

Figure \ref{q4_dn_kn} shows the relation between $d_{min}/N$ and $K/N$ when $q=4$. 
The code length increases exponentially when the dimension increases, but $K/N$ much decreases. 

\subsection{Performance of shortened codes}

\begin{figure}[h]
  \begin{center}
    \includegraphics[width=7.8cm,clip]{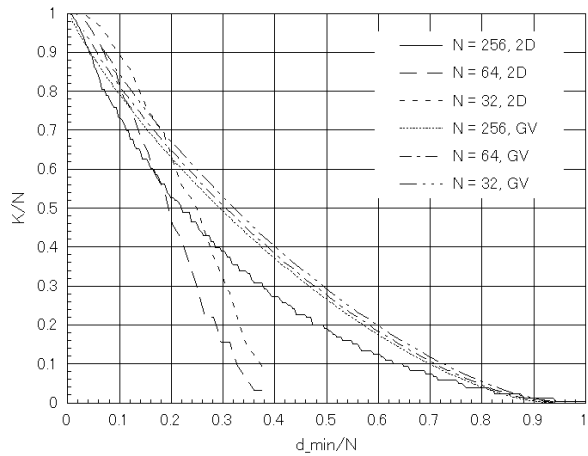}
   \caption[]{Relation between $d_{min}/N$ and $K/N$\ $(2-\textrm{dimensional},\ q=16)$}
    \label{q16_dn_kn}
  \end{center}
\end{figure}

Figure \ref{q16_dn_kn} shows the relation between $d_{min}/N$ and $K/N$ of the shortened 2-dimensional codes when $q=16$. 
Gilbert-Varshamov bounds are also shown in Fig.\ref{q16_dn_kn}. 
When $d_{min}/N$ is small, the shortened codes have higher performance. 
Especially the shortened code of length $N=32$ is beyond the Gilbert-Varshamov bound when $d_{min}/N \leq 0.15$. 

\section{Conclusion}

This paper has proposed a new class of multi-dimensional nonsystematic Reed-Solomon codes that are constructed based on 
the multi-dimensional Fourier transform over a finite field. 
The proposed codes are the extension of the nonsystematic Reed-Solomon codes to multi-dimension. 
The code length of the Reed-Solomon codes can be lengthened by extending the dimension. 
Though the code length increases exponentially when the dimension increases, the code rate decreases. 
The nonsystematic Reed-Solomon codes are the maximum distance separable codes, but the proposed codes are not. 
However there exist some superior shortened 2-dimensional codes that are beyond the Gilbert-Varshamov bound 
when the minimum distance is small. 

The codes presented by Shen, et al., which are constructed using a location set contained in a multidimensional 
affine or projective space over a finite field, seem to be equivalent to the proposed codes.


\begin{thebibliography}{9}

\bibitem{mac}
F.J.MacWilliams and N.J.A.Sloane: "The theory of error-correcting codes," North Holland Publishing Company (1977).

\bibitem{blahut}
R.E.Blahut: "Theory and practice of error control codes," Addison-Wesley Publishing Company (1983). 

\bibitem{reed}
I.S.Reed and G.Solomon, "Polynomial codes over certain finite fields," J.SIAM, vol.8, pp.300-304 (1960).

\bibitem{shiozaki1}
A.Shiozaki: "New class of codes based on two-dimensional Fourier transforms over finite fields," Electronics Letters, Vol.30, No.22, pp.1832-1833 (1994).

\bibitem{shiozaki2}
A.Shiozaki: "A new class of error-correcting-codes based on multi-dimensional Fourier transform over finite field," Proc. of the 17th Symposium on Information Theory 
and Its Applications (SITA '94), pp.225-228 (Hiroshima, Japan, Dec.6-9, 1994). (in Japanese)

\bibitem{shen}
B.Z.Shen and K.K.Tzeng: "Multidimensional extension of Reed-Solomon codes," Proc. of 1998 IEEE International Symposium on Information Theory, p.54 
(Cambridge, MA, USA, Aug.16-21, 1998). 

\end{thebibliography}
\end{document}